\documentstyle[12pt]{article}
\newcommand{\sect}[1]{\setcounter{equation}{0}\section{#1}}

\renewcommand{\thesection}{\arabic{section}}
\renewcommand{\thefootnote}{\fnsymbol{footnote}}
\newcommand{\bea}{\begin{eqnarray}}
\newcommand{\ena}{\end{eqnarray}}
\newcommand{\vs}[1]{\vspace{#1 mm}}

\renewcommand{\a}{\alpha}
\renewcommand{\b}{\beta}
\renewcommand{\c}{\gamma}

\newcommand{\e}{\epsilon}
\newcommand{\p}{\pi}
\newcommand{\n}{\nu}
\renewcommand{\t}{\tau}
\newcommand{\z}{\omega}
\newcommand{\G}{\Gamma}

\newcommand{\PR}[1]{Phys.\ Rev.\ {\bf #1}}

\newcommand{\AJ}[1]{Astorophys. \ J.\ {\bf #1}}
\newcommand{\JMP}[1]{J.\ Math.\ Phys.\ {\bf #1}}

\begin{document}
\topmargin 0pt
\oddsidemargin 5mm

\begin{titlepage}
\setcounter{page}{0}

\begin{flushright}
OU-HET 246\\
gr-qc/9605057\\
May,1996
\end{flushright}
\vs{5}
\begin{center}
{\Large{\bf Analytic Solutions of the Regge-Wheeler Equation \\
     and the Post-Minkowskian Expansion}}\\
\vs{8}
{\large  
Shuhei Mano,\footnote{e-mail address: mano@phys.wani.osaka-u.ac.jp}
Hisao Suzuki\footnote{e-mail address: hsuzuki@particle.phys.hokudai.ac.jp}
and Eiichi Takasugi\footnote{e-mail address: 
takasugi@phys.wani.osaka-u.ac.jp}
}\\
\vs{8}
{\em Department of Physics,
Osaka University \\ Toyonaka, Osaka 560, Japan} \\
and \\
{\em Department of Physics,
Hokkaido  University \\  Sapporo 060, Japan\dag} \\
\end{center}
\vs{8}
\centerline{{\bf Abstract}}
  Analytic solutions of the Regge-Wheeler equation  
are presented in the form of series of hypergeometric functions 
and Coulomb wave functions which have different regions of 
convergence. Relations between these solutions 
are established. 
The  series solutions are given as the Post-Minkowskian expansion 
with respect to a parameter 
$\e \equiv 2M\z$,   
$M$   being the mass of black hole. This expansion 
corresponds to the 
post-Newtonian expansion when they are applied to 
the gravitational radiation 
from a particle in circular orbit around a black hole. 
These solutions can also be  useful  for  numerical computations.   
\end{titlepage}

\newpage
\renewcommand{\thefootnote}{\arabic{footnote}}
\setcounter{footnote}{0}

\sect{Introduction}
\indent
In our previous work\cite{MST}, we presented analytic solutions of 
Regge-Wheeler (RW) equation in the form of series of hypergeometric 
functions. We proved that recurrence 
relations among hypergeometric functions as given in Appendix A in this text 
and showed that coefficients of series are systematically determined 
in the power series of $\e \equiv 2M\z$, $M$   being the mass of black hole. 
We also presented analytic solutions 
in the form of series of Coulomb wave functions which turn out to be 
the same as those given by Leaver\cite{Leaver}. We found that 
the  series of solutions are characterized by the renormalized angular 
momentum which turns out to be identical. Then, we obtained a 
good solution by matching these two types of solutions.

This method can be extended for Teukolsky equation\cite{Teukolsky} 
 in Kerr geometry. 
In this case, the coefficients of series of hypergeometric functions and 
also those of series of Coulomb wave functions series satisfy 
the three term recurrence relations.  About these recurrence relations, 
Otchik\cite{Otchik} made an important observation that  
the recurrence relation for both series are identical, which 
 enabled to relate these two solutions\footnote{In his paper, the 
relation between the series of hypergeometric functions and the 
series of Coulomb wave functions are related in the intermediate 
region where both series converge, though the series which he treated 
are not the solutions of Teukolsky equation.}.  Following the discussion by  
Otchik\cite{Otchik}, Mano, Suzuki and Takasugi\cite{MST2} 
  extended our analysis to  
Teukolsky equation in Kerr geometry 
and reported analytic solutions. We discussed the convergence 
regions of these series and the relation between two solutions of 
difference regions of convergences. The series are expressed in the 
$\e$ expansion which corresponds to the Post-Minkowskian expansion and 
also to the post-Newtonian expansion when they are applied to 
the gravitational radiation 
from a particle in circular orbit around a black hole.  

In this paper, we present analytic solutions of RW equation and discuss 
the analytic properties of these solutions by reorganizing our previous 
work\cite{MST} following our work on Teukolsky equation\cite{MST2}.  
Solutions  are given  in the form of series of hypergeometric functions 
and Coulomb wave functions and the convergence of these series are studied.  
The relation of these solutions with different regions of convergence 
are discussed. The $\e$ expansions of solutions are given up to the 
second order. It is not difficult to obtain solutions in the expansion of 
$\e$ which are used to estimate physical quantities in the gravitational 
wave astrophysics as discussed by Poisson and Sasaki\cite{PS}. 
It is interesting to compare solutions given in $\e$ expansion and 
numerical solutions to test the accuracy of the $\e$ expansion. 
Since the convergence of series is fast, solutions can also be used for 
numerical computations. 

The Regge-Wheeler equation is given by\cite{TN} 
\bea
X''+\left[{{1\over{z-\e}}-{1\over{z}}} \right]X'+\left[1+{2\e\over{z-\e}}+
{\e^2\over{(z-\e)^2}}-{l(l+1)\over{z(z-\e)}}+
{3\e\over{z^2(z-\e)}} \right]X&=&0,\nonumber \\
&&
\ena
where $\e=2M\z,z=\z r$, $M$ being the mass of Schwarzschild black hole and 
$\e$ the angular frequency. In the following, we summarize our result.
 
In Sec.II, we derive  analytic solutions of RW equation in the 
form of series of hypergeometric functions. They are  
 $X_{in}^{\n}$ and  $X_{out}^{\n}$ which satisfy the 
incoming boundary condition and the outgoing boundary condition 
on the horizon. These solutions are labeled by the renormalized 
angular momentum $\n=l+O(\e)$ which is determined by requiring 
the convergence of the series. It can be shown that $-\n-1$ is also 
the renormalized angular momentum. Solutions $X_{in}^{\n}$ and  
$X_{out}^{\n}$ are invariant under the exchange of $\n$ and 
$-\n-1$. These solutions can be   
expressed as combinations of other type of independent solutions 
$X_{0}^{\n}$ and  $X_{0}^{-\n-1}$.   These series solutions are 
convergent except for infinity as shown in Appendix B. 

In Sec.III,  solutions are given in the form of 
series of Coulomb wave functions. They are 
$X_C^{\n}$ and $X_C^{-\n-1}$. They are shown to be convergent 
for $z > \e$ as shown in Appendix B. These solutions  can be expressed by 
linear combination 
of solutions which satisfy the incoming and outgoing boundary conditions, 
$X_{Cin}^{\n}$ and $X_{Cout}^{\n}$. Because the three term recurrence 
relation among coefficients is the same of the hypergeometric function 
series, the renormalized angular momentum for the Coulomb wave function 
series is the same as for the hypergeometric case.
 
In Sec.IV, we show that $X_C^{\n}$  is proportional to  $X_{0}^{\n}$. 
This is due to the fact that  
both series are specified  by the same renormalized angular momentum $\n$. 
The same is true for and $X_C^{-\n-1}$ and   $X_{0}^{-\n-1}$. 
Now we have  analytic solutions both of which is expressed in the form of 
series of 
hypergeometric functions which is convergent except infinity and 
is  expressed in the form of Coulomb wave functions which is convergent 
for $z > \e$. 

In Sec.V, we discuss the relation between solutions of RW equation and 
solutions of Teukolsky equation. For some solutions, the relations can 
be explicitly proved.

In Sec.VI, the Post-Minkowski ($\e$) expansion of solutions are given 
up to the second order of $\e$. The renormalized angular momentum $\n$ 
is expressed as a form of $\e$ expansion. We present the second 
order result.  

Summary and discussions are given in Sec.VII and the derivation of 
the three term recurrence relations among hypergeometric functions 
and among Coulomb wave functions which are crucial to derive the 
three term recurrence relation among coefficients are given.

\indent

\sect{Analytic solution in the form of series of hypergeometric functions}
\indent
 
The RW equation has two regular singularities at $r=0, 2M$ 
and an irregular singularity at $r=\infty$. 
In order to obtain the solution 
in the form of series of hypergeometric functions, we have to deal with 
these regular singularities.  In particular, we eliminate the terms of 
$1/z^2$ and $1/(z-\e)^2$ in Eq.(1.1) by taking the parameterization 
\bea
X_{in}^{\n}&=&e^{i\e(x-1)}(-x)^{-i\e}
(1-x)^{-1}p_{in}^{\n}(x),
\ena
where  $x=1-r/2M=1-z/\e$ and $\e=2M\z$. 
Then, the RW equation becomes
\bea
&&x(1-x){p_{in}^{\n}}''+[1-2i\e+2(1+i\e)x]{p_{in}^{\n}}'
+(l-1-i\e)(l+2+i\e){p_{in}^{\n}}\nonumber\\
&=&-i\e[2x(1-x){p_{in}^{\n}}'+2xp_{in}^{\n}+(1+i\e){p_{in}^{\n}}],
\ena
We obtain the solution of this equation in the form of series of 
hypergeometric functions. For this, we introduce a parameter $\n$ 
and seek the solution in the form 
\bea
p_{in}^{\n}(x)=\sum_{n=-\infty}^{\infty}a_{n}^{\n}p_{n+\n}(x),
\ena
where 
\bea
p_{n+\n}(x)
&=&{\G(n+\n-1-i\e)\G(-n-\n-2-i\e)\over{\G(1-2i\e)}}\nonumber\\
&\times&F(n+\n-1-i\e,-n-\n-2-i\e;1-2i\e;x).
\ena
As an initial condition, we set that $a_n^{\n}=1$ for $n=0$ and $0$ for 
others so that $\n=l +O(\e)$. 

In order for the coefficients of  series (2.3) to be a solved, 
it is essential that 
 the coefficients $a_n^{\n}$'s satisfy the three term recurrence relation. 
For this, it is essential that terms such as $x(1-x){p'}_{n+\n}$ 
and $xp_{n+\n}$ are  expressed as linear combinations of $p_{n+\n+1}$, 
$p_{n+\n}$ and $p_{n+\n-1}$. This turns out to be  true and the recurrence 
relations and the derivation of these relations are given in Appendix A.  

By substituting 
the form in Eq.(2.3) into the radial  RW equation (2.2) and using the 
recurrence relations in Eqs.(A.7) and (A.9) in Appendix A, 
we find that $p_{in}^{\n}$ becomes a solution 
if the following  recurrence relation among the coefficients is 
satisfied:
\bea
\a_n^{\n} a_{n+1}^{\n}+\b_n^{\n} a_n^{\n}+\c_n^{\n} a_{n-1}^{\n}=0,
\ena
where
\bea
\a_n^{\n}&=&-{i\e(n+\n-1+i\e)(n+\n-1-i\e)(n+\n+1-i\e)
\over{(n+\n+1)(2n+2\n+3)}},
\ena
\bea
\b_n^{\n}&=&(n+\n)(n+\n+1)-l(l+1)+2\e^2
+{\e^2(4+\e^2) \over{(n+\n)(n+\n+1)}},
\ena
\bea
\c_n^{\n}&=&{i\e(n+\n+2+i\e)(n+\n+2-i\e)(n+\n+i\e)
\over{(n+\n)(2n+2\n-1)}}.
\ena

By introducing the continued fractions 
\bea
R_n (\n)={a_n^{\n}\over{a_{n-1}^{\n}}},
\qquad
L_n(\n)={a_n^{\n}\over{a_{n+1}^{\n}}},
\ena
we find 
\bea
R_n(\n)=-{\c_n^{\n}\over{\b_n^{\n}+\a_n^{\n}R_{n+1}(\n)}},
\qquad
L_n(\n)=-{\a_n^{\n}\over{\b_n^{\n}+\c_n^{\n}L_{n-1}(\n)}}.
\ena
From these equations, we can evaluate  the coefficients by taking 
the initial condition $a_0^{\n}=1$. The parameter $\n$ is determined 
by requiring that  
the coefficients obtained by using $R_n(\n)$ agree with 
those by using $L_n(\n)$. This is the condition of the convergence 
of the series. This condition gives  the transcendental equation for $\n$ 
\bea
R_n(\n)L_{n-1}(\n)=1.
\ena
The solution $\n$ of this equation is called the renormalized 
angular momentum. By using this $\n$, 
 the series is shown to 
be convergent in all the complex plane of $x$ except for $x=\infty$ 
as discussed  in Appendix B. 

The normalization of $X_{in}^{\n}$  is given on the  horizon
 $(x = \e)$  as
\bea
X_{in}^{\n}\rightarrow \left (e^{-i\e}
\sum_{n=-\infty}^{\infty}a_n^{\n}\frac {\G(n+\n-1-i\e)\G(-n-\n-2-i\e)}
{\G(1-2i\e)}\right ) e^{-i\e \ln (-x)}.
\ena
 
Now we can show that $-\n-1$  satisfies  
 Eq.(2.13) so that $-\n-1$ 
is also the renormalized angular momentum\cite{MST2}. 
As for the recurrence relation (2.8), we find   
$\alpha_{-n}^{-\n-1}=\gamma_n^{\n}$ and   
$\gamma_{-n}^{-\n-1}=\alpha_n^{\n}$ so that  
 $a_{-n}^{-\n-1}$ satisfies the same recurrence relation as $a_{n}^{\n}$ does. 
Thus  if we choose 
$a_0^{\n}=a_0^{-\n-1}=1$,  we have 
\bea
a_n^{\n}=a_{-n}^{-\n-1}.
\ena

  By using the 
formula
\bea
p_{n+\n}(x)&=&\frac{\G (-n-\n-2-i\e)\G (2n+2\n+1)}
{\G (n+\n+3-i\e)}(1-x)^{n+\n+2+i\e}\nonumber\\
&\times&F(-n-\n-2-i\e,-n-\n+2-i\e;-2n-2\n;\frac{1}{1-x})\nonumber\\
&+&\frac{\G (n+\n-1-i\e)\G (-2n-2\n-1)}
{\G (-n-\n+2-i\e)}(1-x)^{-n-\n+1+i\e}\nonumber\\
&\times&F(n+\n-1-i\e,n+\n+3-i\e;2n+2\n+2;\frac{1}{1-x}),
\ena
we can show
\bea
X_{in}^{\n}=X_0^{\n}+X_0^{-\n-1},
\ena
where 
 \bea
X_0^{\n}&=&e^{i\e(x-1)}(-x)^{-i\e}
(1-x)^{\n+1+i\e}\sum_{n=-\infty}^{\infty}
a_n^{\n}{\G(-n-\n-2-i\e)\G(2n+2\n+1)
\over{\G(n+\n+3-i\e)}}\nonumber\\
&\times&(1-x)^{n}F(-n-\n-2-i\e,-n-\n+2-i\e;-2n-2\n;{1\over{1-x}}).
\ena

We consider the solution which satisfies the outgoing boundary 
condition on the  horizon. The outgoing solution 
is parametralizd by  
\bea
X_{out}^{\n}=e^{i\e(x-1)}(-x)^{i\e}
(1-x)^{-1}p_{out}^{\n}.
\ena
Then $p_{out}^{\n}$ should satisfy the following equation,
\bea
&&x(1-x){p_{out}^{\n}}''+[1+2i\e+2(1-i\e)x]{p_{out}^{\n}}'
+(l-1+i\e)(l+2-i\e){p_{out}^{\n}}\nonumber\\
&=&-i\e[2x(1-x){p_{out}^{\n}}'+2(1-2i\e)xp_{out}^{\n}+(1+5i\e){p_{out}^{\n}}],
\ena
Now we expand $p_{out}^{\n}$ as 
\bea
p_{out}^{\n}(x)=\sum_{n=-\infty}^{\infty}\tilde{a_n^{\n}}\tilde{p}_{n+\n}(x),
\ena
where 
\bea
\tilde{p}_{n+\n}(x)&=&
{\G(n+\n-1+i\e)\G(-n-\n-2+i\e)\over{\G(1+2i\e)}}\nonumber\\
&\times&F(n+\n-1+i\e,-n-\n-2+i\e;1+2i\e;x).
\ena
Similarly to the solution satisfying the incoming boundary condition,  
we find that the above series becomes a solution if $\tilde{a}_n^{\n}$ 
satisfies the same  three term 
recurrence relation as $a_n^{\n}$ does. 
Then, by choosing  $\tilde{a}_0^{\n}=\tilde{a}_n^{-\n-1}=1$, we find 
\bea
\tilde{a}_0^{\n}=a_0^{\n}.
\ena 
By using Eq.(2.21),   the outgoing solution is given by 
\bea
X_{out}^{\n}&=&e^{i\e(x-1)}(-x)^{i\e}
(1-x)^{-1}\sum_{n=-\infty}^{\infty}a_n^{\n}\tilde{p}_{n+\n}(x)
\nonumber\\
&=&A_{\n}X_0^{\n}+A_{-\n-1}X_0^{-\n-1},
\ena
where
\bea
A_{\n}={\sin\p(\n+i\e)\over{\sin\p(\n-i\e)}}.
\ena
This relation also shows that $X_0^{\n}$ and $X_0^{-\n-1}$ are 
independent solutions of Eq.(2.1). 

\sect{Analytic solutions in the form of series of Coulomb wave functions}  
\indent

 Analytic solution in the form of series of  Coulomb wave functions 
are given by Leaver\cite{Leaver}. Here, we parameterization to 
remove the singularity at $r=2M$.  By using 
a variable $z=\z r =\e(1-x)$, we take the following form
\bea
X_C^{\n}&=&\left(1-{\e \over{z}}\right)^{-i\e}f_{\n}(z).
\ena
Then, we find 
\bea
&&z^2f_{\n}''+[z^2+2\e z-l(l+1)]f_{\n} \nonumber \\
&=&\e[ z(f_{\n}''+f_{\n})-(1-2i\e)f_{\n}'
-\frac 1{z}(3+2i\e+\e^2 )f_{\n}-\e f_{\n}].
\ena
If we consider $\n$ to be $\n=l+O(\e)$, the right-hand side of 
Eq.(3.2) is the quantity of order $\e$ so that this 
equation is a suitable one to obtain the solution in the 
expansion of $\e$. 

Here we shall obtain the exact solution by expanding  $f_{\n}(z)$ 
in terms of Coulomb functions with the renormalized angular momentum 
$\n$,
\bea
f_{\n}=\sum_{n=-\infty}^{\infty}i^n 
\frac{\G(n+\n-1-i\e)\G(n+\n+1-i\e)}{\G(n+\n+1+i\e)\G(n+\n+3+i\e)}
b_n^{\n}F_{\n+n}(z),
\ena
where $F_{n+\n}$ is the unnormalized Coulomb wave function, 
\bea
F_{n+\n}=e^{-iz}(2z)^{n+\n}z
{\Gamma(n+\n+1+i\e)\over{\Gamma(2n+2\n+2)}}
\Phi(n+\n+1+i\e,2n+2\n+2;2iz).
\ena  
  
By substituting Eq.(3.3) into Eq.(3.2) and using the 
recurrence relations in Eqs.(A.18) and (A.19) in Appendix A, 
we find that the coefficients $b_n^{\n}$ satisfy the same three term 
recurrence relation as  $a_n^{\n}$ does. Now  we choose 
$b_0^{\n}=b_0^{-n-1}=1$ and then 
\bea
b_n^{\n}=a_n^{\n}.
\ena  
The fact that  the recurrence relation  for the Coulomb expansion 
case is the same to  the one for the hypergeometric case means that   
the renormalized angular momenta $\n$ are the same for   both 
solutions. This is important for these two solutions to be related 
each other. Therefore, $-\n-1$ is also the renormalized angular 
momentum. As a result, another independent solution is obtained by 
replacing $\n$ with $-\n-1$. Thus two independent solutions are 
$X_C^{\n}$ and $X_C^{-\n-1}$ and $X_C^{\n}$ is given by 
\bea
X_C^{\n}&=&\left(1-{\e \over{z}}\right)^{-i\e}
\sum_{n=-\infty}^{\infty}i^n 
\frac{\G(n+\n-1-i\e)\G(n+\n+1-i\e)}{\G(n+\n+1+i\e)\G(n+\n+3+i\e)}
a_n^{\n}F_{\n+n}(z).
\ena

Another forms of these solutions suitable for examining  
the asymptotic behavior 
are  
\bea
X_C^{\n}=X_{C\,in}^{\n}+X_{C\,out}^{\n},
\ena
\bea
X_C^{-\n-1}&=&X_{C in}^{-\n-1}+X_{C out}^{-\n-1}\nonumber\\
      &=&ie^{-i\pi (\n+1)}\frac{\sin \pi(\n+i\e)}{\sin \pi(\n-i\e)}
X_{C in}^{\n}
-ie^{i\pi (\n+1)}X_{C out}^{\n},
\ena
where
\bea
X_{C\,out}^{\n}&=&e^{iz}z^{\n+1}\left(1-{\e \over{z}}\right)
^{-i\e}2^{\n}e^{-\pi \e}e^{-i\p(\n+1)}\nonumber\\
&\times&\sum_{n=-\infty}^{\infty}i^n\frac{\G(n+\n-1-i\e)\G(n+\n+1-i\e)}
{\G(n+\n+1+i\e)\G(n+\n+3+i\e)}a_n^{\n}(-2z)^n\nonumber\\
&\times&\Psi(n+\n+1-i\e,2n+2\n+2;-2iz),
\ena
\bea
X_{C\,in}^{\n}&&=e^{-iz}z^{\n+1}\left(1-{\e \over{z}}
\right)^{-i\e}2^{\n}e^{-\pi \e}e^{i\p(\n+1)}\nonumber\\
&\times&\sum_{n=-\infty}^{\infty}i^n {\G(n+\n-1-i\e)\over{\G(n+\n+3+i\e)}}
a_n^{\n}(-2z)^n
\Psi(n+\n+1+i\e,2n+2\n+2;2iz).\nonumber\\
\ena
 
In the end of this section, we present another set of solutions as 
\bea
\tilde{X}_C^{\n}&=&\left(1-{\e \over{z}}\right)^{i\e} 
  \sum_{n=-\infty}^{\infty}i^n 
\frac{\G(n+\n-1+i\e)\G(n+\n+1-i\e)}{\G(n+\n+1+i\e)\G(n+\n+3-i\e)}
     a_n^{\n}F_{\n+n}(z),
\ena
where $F_{\n+n}$ is defined in Eq.(3.4).  
In this parameterization, $\tilde{X}_C^{\n}$ and $\tilde{X}_C^{-\n-1}$ 
are independent solutions. There are two sets of independent solutions, 
$\{X_C^{\n},X_C^{-\n-1}\}$ and $\{\tilde{X}_C^{\n},\tilde{X}_C^{-\n-1}\}$. 
In the following, we consider a set $\{X_C^{\n},X_C^{-\n-1}\}$ which are 
related directly to a set of hypergeometric type solutions 
$\{X_0^{\n},X_0^{-\n-1}\}$ given in the proceeding section.

\sect{The relation between two solutions}

First we notice that $R_0^{\n}$ and $R_C^{\n}$ are solutions of Teukolsky 
equation. Second we see that if we expand  
 these solutions in  Laurent series of $1-x=z/\e \kappa $,  
both solutions give the series with   
the same characteristic exponent at 
$x \rightarrow \infty $. Thus, $R_0^{\n}$ must be proportional 
to  $R_C^{\n}$,
\bea
X_0^{\n}=K_{\n}X_C^{\n}.
\ena
The constant factor  $K_{\n}$ is determined  
 by comparing  like terms of these  series. 
We find 
\bea
K_{\n}&=&-{\p i^r 2^{-\n-r} \e^{-\n-r-1}
\over{\G(1+r+\n+i\e)\G(-1+r+\n+i\e)
\G(3+r+\n+i\e)}\sin\p(\n+i\e)}\nonumber\\
&\times&\sum_{n=r}^{\infty}
{\G(n+\n-1+i\e)\G(n+r+2\n+1)\over{(n-r)!\G(n+\n+3-i\e)}}a_n^{\n}\nonumber\\
&\times&\left [{\sum_{n=-\infty}^{r}
{{\G(n+\n-1-i\e)\G(n+\n+1-i\e)a_n^{\n}}\over{
\G(n+\n+1+i\e)\G(n+\n+3+i\e)(r-n)!\G(n+r+2\n+2)}}}\right]^{-1},\nonumber\\
\ena
where $r$ is an arbitrary integer.

By using these relations,  $R_{in}^{\n}$ can be written by 
using the Coulomb expansion solutions as
\bea
X_{in}^{\n}&=&(K_{\n}X_{C\,in}^{\n}+K_{-\n-1}X_{C\,in}^{-\n-1})
+(K_{\n}X_{C\,out}^{\n}+K_{-\n-1}X_{C\,out}^{-\n-1})\nonumber\\
&=&\left [ K_{\n}+ie^{-i\pi (\n+1)}\frac{\sin \pi(\n+i\e)}
{\sin \pi(\n-i\e)}K_{-\n-1}\right ]
X_{C in}^{\n} +(K_{\n}-
ie^{i\pi (\n+1)}K_{-\n-1})X_{C out}^{\n}.
\nonumber\\
\ena
The asymptotic behavior at $z \rightarrow \infty $ is 
\bea
X_{in}^{\n}&=&A_{out}^{\n}e^{iz}z^{i\e}+A_{in}^{\n}
e^{-iz}z^{-i\e},
\ena
where $A_{out}^{\n}$ and $A_{in}^{\n}$ are amplitudes of the 
outgoing and incoming waves at infinity  of  the solution which satisfy the 
incoming boundary condition at the outer horizon. They are given by 
\bea
A_{out}^{\n}&=&e^{-{\p\over{2}}\e}2^{-1+i\e}
\left[K_{\n}(-i)^{\n+1}+K_{-\n-1}i^{\n}\right ]
\sum_{n=-\infty}^{\infty}b_n^{\n}(-i)^n,
\ena
and
\bea
A_{in}^{\n}&=&e^{-{\p\over{2}}\e}2^{-1-i\e}\left
[K_{\n}i^{\n+1}+K_{-\n-1}(-i)^{\n}
{\sin\p(\n+i\e)\over{\sin\p(\n-i\e)}}\right]
\nonumber\\
&\times&\sum_{n=-\infty}^{\infty}b_n^{\n}i^n{\G(n+\n+1+i\e)
\over{\G(n+\n+1-i\e)}}.
\ena

One application of these amplitudes is to derive the  absorption 
coefficients. 
The absorption 
coefficient $\G$ can be expressed in terms of $A^{in}$ and $A^{out}$ 
as follows;
\bea
\G^{\n}=1-\left|{A_{out}^{\n}\over{A_{in}^{\n}}}\right|^2
\ena

The upgoing solution which satisfy the outgoing boundary condition 
at infinity is given by $X_{C out}^{\n}$. This solution is expressed 
in terms of $X_0^{\n}$ and   $X_0^{-\n-1}$ as follows

\bea
X_{up}^{\n}&\equiv &\frac{\G(\n-1-i\e)\G(\n+1-i\e)}{\G(\n+1+i\e)\G(\n+3+i\e)}
X_{C out}^{\n}\nonumber\\
 &=& \left [ \frac{\sin \pi(\n+i\e)}
{\sin \pi(\n-i\e)}(K_{\n})^{-1}X_0^{\n}-ie^{i\pi \n}
(K_{-\n-1})^{-1}X_0^{-\n-1}   \right ]\nonumber\\
 && \hskip 2mm \times \left [ e^{2i\pi \n}+\frac{\sin \pi(\n+i\e)}
{\sin \pi(\n-i\e)}  \right ]^{-1}.
\ena

\sect{Relation between solutions of Teukolsky  and Regge-Wheeler 
      equations}

Let us present some solutions of Teukolsky equation in Ref.1 
for spin $s=-2$
in the Schwarzschild limit ($a \to 0$) :

The  solution which satisfies the incoming boundary condition
 on the horizon is expressed as a series of hypergeometric functions as
\bea
R_{in(-2)}^{\n}&=&e^{i\e x}(-x)^{2-i\e}
\sum_{n=-\infty}^{\infty}a_{n}^{\n,T}
 F(n+\n+1-i\t,-n-\n-i\t;3-2i\e;x)
\ena    
and coefficients $a_{n}^{\n,T} $  are determined by 
the three term recurrence relation, 
\bea
\a_n^{\n,T} a_{n+1}^{\n,T}+\b_n^{\n} a_{n}^{\n,T}
+\c_n^{\n,T} a_{n-1}^{\n,T}=0,
\ena
\bea
\a_n^{\n,T}&=&{i\e (n+\n-1+i\e)(n+\n-1-i\e)(n+\n+1+i\e)
\over{(n+\n+1)(2n+2\n+3)}},
\ena
\bea
\c_n^{\n,T}&=&-{i\e (n+\n+2+i\e)(n+\n+2-i\e)(n+\n-i\e)\over{(n+\n)(2n+2\n-1)}},
\ena
and most importantly  $\b_n^{\n}$ is the same as the one in Eq.(2.9). 
By comparing the recurrence relation in Eq.(2.5) and (5.4), we find with 
the condition with $a_0^{\n,T}=a_0^{-\n-1,T}=1$ that 
\bea
a_{n}^{\n,T}=(-1)^n\frac{\G(\n+1+i\e)\G(n+\n+1-i\e)}
{\G(\n+1-i\e)\G(n+\n+1+i\e)} a_{n}^{\n}.
\ena

A solution valid at infinity is given by 
\bea
R_{C(-2)}^{\n}&=&z\left(1-{\e \over{z}}\right)^{i\e} 
  \sum_{n=-\infty}^{\infty}(-i)^n \nonumber\\
&&\hskip 3mm \times
\frac{\G(n+\n-1+i\e)\G(n+\n-1-i\e)\G(n+\n+1+i\e)}
{\G(n+\n+3+i\e)\G(n+\n+3-i\e)\G(n+\n+1-i\e)} a_{n}^{\n,T}
F_{\n+n}^T(z),
\ena
where 
\bea
F_{n+\n}^T=e^{-iz}(2z)^{n+\n}z
{\Gamma(n+\n+3+i\e)\over{\Gamma(2n+2\n+2)}}
\Phi(n+\n+3+i\e,2n+2\n+2;2iz).
\nonumber\\
\ena  
Another solution is $R_{C(-2)}^{-\n-1}$. 
These solutions are given in the expansion around the origin and are  
different from the ones given in our previous paper\cite{MST} which 
corresponds to solutions  expanded around the  horizon. 
 
Let us first consider the following relation between a solution $R$ 
of the Teukolsky 
equation  for $s=-2$ and a solution $X$ of the RW equation,
\bea
R&=&\Delta\left({d\over{dr^{*}}}+i\z\right){r^2\over{\Delta}}
\left({d\over{dr^{*}}}+i\z\right)rX(z),
\ena
where $r^{*}=r+2M\ln(r/2M-1)$ and $\Delta =r^2-2Mr$. 
We  substitute the incoming solution $X_{in}^{\n}$ in Eq.(2.1) 
for $X$ in Eq.(5.1) and find 
\bea
R&=&\frac{\e}{\omega}(-x)^2(\frac{d}{dx}-i\e +\frac{i\e}{x})^2 
  e^{i\e (x-1)}(-x)^{-i\e}p_{in}^{\n}\nonumber\\
&=&\frac{\e}{\omega}e^{i\e (x-1)}(-x)^{2-i\e}\frac{d^2}{dx^2}
p_{in}^{\n}
\nonumber\\
&=&\frac{\e}{\omega}
e^{i\e (x-1)}(-x)^{2-i\e}
\sum_{n=-\infty}^{\infty}
{{\G(n+\n+1-i\e)\G(-n-\n-i\e)}
\over{\G(3-2i\e)}}
a_{n}^{\n}\nonumber\\
&&\hskip 1cm \times F(n+\n+1-i\e,-n-\n-i\e;3-2i\e;x),
\nonumber\\
\ena
where $p_{in}^{\n}$ is defined in Eq.(2.4).  
By substituting  Eq.(5.6) 
into (5.9), we find that  
$R$ in Eq.(5.8) is proportional to   
 $R_{in(-2)}^{\n}$. We find 
\bea
\Delta\left({d\over{dr^{*}}}+i\z\right){r^2\over{\Delta}}
\left({d\over{dr^{*}}}+i\z\right)rX_{in}^{\n}=
e^{-i\e}\frac{\e}{\omega}\frac{\G(\n+1-i\e)\G(-\n-i\e)}{\G(3-2i\e)}
R_{in(-2)}^{\n}.
\ena
  
We next consider the inverse relation 
\bea
X&=&\frac{r^5}{c_0 \Delta}\left({d\over{dr^{*}}}-i\z\right){r^2\over{\Delta}}
\left({d\over{dr^{*}}}-i\z\right)\frac{R}{r^2}.
\ena
For this relation, we consider a solution of the Teukolsky equation given 
in Eq.(5.4). 
Now, we substitute $R_{C(-2)}$ into Eq.(5.7) and find 
\bea
X&=&\frac{\omega}{c_0}z^3(\frac{d}{dz}-i-\frac{i\e}{z-\e})^2
\left ( \frac{R_{C(-2)}^{\n}}{z^2}\right )
\nonumber\\
 &=&\frac{\omega}{c_0}z^3(\frac{d}{dz}-i-\frac{i\e}{z-\e})^2
  e^{iz}(z-\e)^{i\e}[e^{-iz}(z-\e)^{-i\e}z^{-2}R_{C(-2)}^{\n}]
\nonumber\\
 &=&\frac{\omega}{c_0}\frac{\G(\n+1+i\e)}{\G(\n+1-i\e)}z^3 e^{iz}(z-\e)^{i\e}  
\sum_{n=-\infty}^{\infty}i^n
 {{\G(n+\n-1+i\e)\G(n+\n-1-i\e) }
\over{\G(n+\n+3-i\e) \Gamma(2n+2\n+2)}}a_n^{\n}
\nonumber\\
&&\hskip 5mm \times \frac{d^2}{dx^2} e^{-2iz}z^{-1-i\e}(2z)^{n+\n}z
\Phi(n+\n+3+i\e,2n+2\n+2;2iz)\nonumber\\
&=&\frac{\omega}{c_0}\frac{\G(\n+1+i\e)}{\G(\n+1-i\e)}(1-\frac{\e}{z})^{i\e}
\sum_{n=-\infty}^{\infty} i^n
\frac{\G(n+\n-1+i\e) \G(n+\n+1-i\e)}
{\G(n+\n+1+i\e)\G(n+\n+3-i\e) }a_n^{\n}F_{n+\n},
\nonumber\\
\ena
where $F_{n+\n}$ is the Coulomb wave function defined in Eq.(3.4). 
The third equality is obtained by substituting $a_n^{\n,T}$ in Eq.(5.4) and 
the final equality is derived by using   the following relation
\bea
\frac{d^2}{dx^2}[e^{-x}x^{c-a+1} \Phi(a,c;x)]=(c-a)(c-a+1)e^{-x}
   x^{c-a-1}\Phi(a-2,c;x).
\ena
By comparing Eq.(5.16) and Eq.(3.21), we find $X$ is equal to 
$\tilde{X}_C^{\n}$ up to the numerical factor. 
 Thus, we obtain the relation 
\bea
 \frac{r^5}{c_0 \Delta}\left({d\over{dr^{*}}}-i\z\right)
{r^2\over{\Delta}}
\left({d\over{dr^{*}}}-i\z\right)\frac{R_{C(-2)}^{\n}}{r^2}=
\frac{\omega}{c_0}\frac{\G(\n+1+i\e)}{\G(\n+1-i\e)}{\tilde X}_C^{\n}.
\ena
 
Equations (5.12) and (5.20) show the relations between a solution of 
Teukolsky equation with $s=-2$ and a solution of the RW equation.

\sect{Post-Minkowskian expansions of solutions}
 In this section, we discuss how to derive the solution in the expansion 
of the small parameter $\e=2M\z$. 
In order to find the solution in Eqs.(2.1) and (4.3) up to some 
power of $\e$, we have to calculate $\n$ and $a_n^{\n}$ to that 
order by using Eq.(2.12) and (2.13) with the condition (2.14) and 
$a_0^{\n}=a_0^{-\n-1}=1$. Other coefficients $b_n^{\n}$ can be calculated 
from $a_n^{\n}$ by using the formula (3.10).

For $a_n^{\n}$ with $n \ge 1$, the equation for $R_n(\n)$ is 
useful. Since   $\a_n^{\n}, \gamma_n^{\n}  \sim O(\e)$ and 
and $\b_n^{\n} \simeq n(n+2l+1) \sim O(1)$, we find 
\bea
R_n(\n) \sim O(\e)
\ena
 for all positive integer $n$. 
As a result with $a_0^{\n}=1$, 
we find  
\bea
a_n^{\n} \sim O(\e^n)\hskip 1cm {\rm for} \hskip 5mm 
n\ge 1.
\ena

Before discussing the coefficients for $n<0$, 
 we  derive the renormalized angular 
momentum $\n$ up to $O(\e^2)$. For this, it is convenient to  use 
the constraint for $n=1$, $R_1(\n)L_0(\n)=1$. We notice that 
$R_1(\n)\sim O(\e)$ so that  $L_0(\n)$ must behave as $O(1/\e)$.  
which requires that $\b_0^{\n} + \gamma_0^{\n}L_{-1}(\n) \sim O(\e^2)$ 
because $\a_0^{\n} \sim O(\e)$. In order to obtain 
$\n$ up to $O(\e)$, we need to know 
 the information of $\b_0^{\n}$ up to $O(\e^2)$ where 
the second order term of $\n$ involves. Thus, we need the 
information about $R_1(\n)$, $L_{-1}(\n)$, $\a_{0}^{\n}$ and 
$\gamma_{0}^{\n}$ up to $O(\e)$. Here we assume that 
$L_{-2}(\n)\sim O(\e)$ whose validity will be discussed later. 
In this situation, $R_1(\n)$, $L_{-1}(\n)$, $\a_{0}^{\n}$ and 
$\gamma_{0}^{\n}$ can be calculated immediately.  
By substituting these to the constraint equation $R_1(\n)L_0(\n)=1$ 
to find 
\bea
\n&=&l+{1\over{2l+1}}\left[-2+{-4\over{l(l+1)}}
+{(l-1)^2(l+3)^2\over{(2l+1)(2l+2)(2l+3)}}-{(l-2)^2(l+2)^2
\over{(2l-1)(2l)(2l+1)}}\right]\e^2\nonumber\\
&&+O(\e^3).
\ena
In general, we can show that $\n$ is a even function of $\e$. We see 
that the continued fractions $R_n(\n)$ and $L_n(\n)$ are determined 
by $\b_n^{\n}$ and $\a_n^{\n}\gamma_{n+1}^{\n}$. Since these quantities 
are even functions,  the parameters in the transcendental 
equation $R_n(\n)L_{n-1}(\n)=1$ aside from $\n$ are even functions of $\e$ 
which 
concludes that   $\n$  an even function of $\e$. Thus, the next 
correction term in $\n$ enters in the 4-th order term of $\e$.

The fact that the correction term of $\n$ starts from the second order term 
of $\e$ simplifies the calculation of the coefficients up to $O(\e^2)$. 

Now we discuss the coefficients for  negative integer $n$ 
which are 
derived by using  the equation for $L_n(\n)$. 
For large negative value of $\mid n \mid$, 
$L_n(\n) \simeq  i\e /2n$. Most of the negative integer value of 
$n$, $L_n(\n) \sim O(\e)$. There arise some exceptions  
for certain values of $n$ because  
the denominator of $\a_n^{\n}$ vanishes  
at $n=-l-1$  and  
also $\b_n^{\n}$ vanishes at $n=-2l-1$ in the zeroth order of $\e$. 
There is a 
 speciality of Regge-Wheeler solutions that $\a_{-l+1}^{\n} \sim O(\e^3)$.   
Because of these, we find
\bea
L_{-l+1}(\n) \sim O(\e^3)&&,
\nonumber\\
L_{-l-1}(\n) \sim O(1)&&, \nonumber\\
L_{-2l-1}(\n) \sim O(1/\e)&&, \nonumber\\
L_n(\n) \sim O(\e) && {\rm for \hskip 1mm all \hskip 1mm others}.
\ena 
From these observations, we find  
\bea
a_n^{\n} &\sim& O({\e}^{\mid n\mid}), \hskip 5mm {\rm for } \hskip 5mm 
  -1 \ge n \ge -l+2,
\nonumber\\
a_{-l+1}^{\n} &\sim& O(\e^{l+1}),
\nonumber\\
a_{-l}^{\n} &\sim&a_{-l-1}^{\n} \sim O(\e^{l+2})
\nonumber\\
a_n^{\n} &\sim& O(\e^{\mid n\mid +1}), \hskip 5mm {\rm for } 
\hskip 5mm -l-2 \ge n \ge  -2l,\nonumber\\
a_n^{\n} &\sim& O(\e^{\mid n\mid -1}), \hskip 5mm {\rm for } 
\hskip 5mm    -2l-1 \ge n.
\ena
With the above order estimates, we see that how many terms should be 
needed to calculate the coefficients with   the specified accuracy   
of $\e$. 
 
Comming back to $\n$, we assumed that $L_{-2}(\n) \sim O(\e)$ which is 
valid since  $l \ge 2$. .

The coefficients $a_n^{\n}$ and also $b_n^{\n}$ up to $O(\e^2)$ are obtained 
explicitly by
\bea
a_{1}&=&{-i(l+3)^2\over{2(l+1)(2l+1)}}\e+
      {(l+3)^2\over{2(l+1)^2(2l+1)}}\e^2+O(\e^3),
\ena
\bea
a_{2}&=&{-(l+3)^2(l+4)^2\over{4(l+1)(2l+1)(2l+3)^2}}\e^2+O(\e^3),
\ena 
\bea
a_{-1}&=&{-i(l-2)^2\over{2l(2l+1)}}\e -
      {(l-2)^2\over{2l^2(2l+1)}}\e^2+O(\e^3),
\ena
\bea
a_{-2}&=&{-(l-3)^2(l-2)^2\over{4l(2l-1)^2(2l+1)}}\e^2+O(\e^3),
\ena 
 
By using these coefficients, we can evaluate the incoming and the 
outgoing amplitudes in infinity. 
From Eq.(4.2), we find by taking $r=0$ that 
$K_{\n} \sim O(\e^{-l-1})$. On the other hand, the estimate of 
$K_{-\n -1}$ needs some care. By taking into account of the 
singular behaviors of gamma functions and the fact that 
the deviation of $\n$ from $l$ starts from the second order of 
$\e$, we find that $K_{-\n -1} \sim O(\e^{l-1} )$. 
Thus we obtain
\bea
\frac {K_{-\n -1}}{K_{\n }} \sim O(\e^{2l}),
\ena
which concludes that $K_{-\n-1}$ term contributes at most the 
order $\e^4$ ( $l\ge 2$). In the approximation up to $O(\e^2)$, 
we can safely neglect $K_{-\n -1}$ term.

Thus, we get the simple expressions for 
the outgoing and the incoming amplitudes as follows;
\bea
A_{out}^{\n}=B_0^{\n} K_{\n}
\sum_{n=-2}^{2}\frac{\G(n+\n-1-i\e)\G(n+\n+1-i\e)}
{\G(n+\n+1+i\e)\G(n+\n+3+i\e)}a_n^{\n},
\ena
and
\bea
A_{in}^{\n}= (B_0^{\n})^*K_{\n}
\sum_{n=-2}^{2}(-1)^n{\G(n+\n-1-i\e)\over{\G(n+\n+3+i\e)}}a_n^{\n},
\ena
where 
\bea
B_0^{\n}=(-i)^{\n+1}2^{-1+i\e}e^{-\pi\e /2} .
\ena
By substituting the coefficients, we can easily calculate the 
amplitudes up to the order $\e^2$. Since the explicit expressions 
are complicated, we present the amplitudes up to $O(\e)$ 
explicitly. We find 
\bea
A_{out}^{\n}&=&B_0^{\n}K_{\n}\frac{\G(\n-1-i\e)\G(\n+1-i\e)}
{\G(\n+1+i\e)\G(\n+3+i\e)}\nonumber\\
&&\hskip 3mm \left\{ 
1-i\e\frac{(l-1)l(l+3)+(l-2)(l+1)(l+2)}
{2l(l+1)(2l+1)}-\e^2\frac{2}{l^2(l+1)^2}\right. \nonumber\\
&&\hskip 5mm \left. 
-\e^2\frac1{4(2l+1)}\left[
\frac{(l-1)l(l+3)(l+4)}{(l+1)(2l+3)^2}+
\frac{(l-3)(l-2)(l+1)(l+2)}{l(2l-1)^2}\right]\right\},
\nonumber\\
\ena
\bea
A_{in}^{\n}=\frac{(B_0^{\n})^*}{B_0^{\n}}\frac{\G(\n+1+i\e)}{\G(\n+1-i\e)}
A_{out}^{\n}.
\ena
The above result shows that the absorption coefficient $\Gamma$ in 
Eq.(4.7) is zero up to the order of $\e$ for  Schwarzschild black hole.

\sect{Summary and Remarks}

We presented  analytical solutions of Regge-Wheeler equation 
in the form of series of hypergeometric functions and Coulomb wave 
functions. The series are characterized by the renormalized angular 
momentum which turns out to be identical for both series which 
enabled us to relate these two series in the intermediate region 
where both series converge. The relation between solutions of 
Regge-Wheeler equation and those 
of Teukolsky solutions. This shows the consistency between 
present solutions and solutions presented in our previous 
paper\cite{MST2}. 
 
Solutions which we found  will be used for the gravitational wave 
astrophysics,  the gravitational wave emitted from a Schwarzschild 
black hole and also emitted from the particle rotating along a 
circular orbit around a black hole. Since solutions are 
given in simpler forms than those of Teukolsky equation, these 
solutions can be obtained in the $\e$ expansion in much higher 
accuracy so that these solutions can be used to test the $\e$ 
expansion. Solutions will be useful for numerical computations 
because the series converges fast. We expect that our solutions 
 will become a powerful weapon to construct the theoretical 
template towards LIGO and VIRGO projects.

\vskip 4cm
{\Huge Acknowledgment}

We would like to thank to M. Sasaki, M.Shibata and T. Tanaka for comments and 
encouragements. This work is supported in part by 
the Japanese Grant-in-Aid for Scientific Research of
Ministry of Education, Science, Sports and Culture, 
No. 06640396.

\newpage

%Appendix format setting
\setcounter{section}{0}
\renewcommand{\thesection}{\Alph{section}}
\newcommand{\apsc}[1]{\stepcounter{section}
\noindent\setcounter{equation}{0}{\Large{\bf{Appendix\,\thesection:\,{#1}}}}}

\apsc{Proof of the recurrence relations among hypergeometric functions and 
Coulomb wave functions}

\noindent
(a) Proof of the recurrence  relations among hypergeometric functions\\

Let us define
\bea
A_{L,n}=\frac{\G(n+L-1-i\e)\G(n-L-2-i\e)}{\G(n+1-2i\e)n!},
\ena
then 
\bea
p_L=\sum_{n=0}^{\infty} A_{L,n} x^n.
\ena
We parameterize $P_{L+1}$ as
\bea
p_{L+1}&=&\sum_{n=0}^{\infty}\frac{\G(n+L-i\e)\G(n-L-3-i\e)}
{\G(n+1-2i\e)n!}x^n
\nonumber\\
&=&\sum_{n=0}^{\infty}\left[1+\frac{2(L+1)}{n-L-3-i\e}\right] A_{L,n}x^n,
\ena
so that we find 
\bea
\sum_{n=0}^{\infty}\frac{A_{L,n}x^n}{n-L-3-i\e}=\frac 1{2(L+1)}(p_{L+1}-p_L).
\ena
Similarly, we find
\bea
\sum_{n=0}^{\infty}\frac{A_{L,n}x^n}{n+L-2-i\e}=\frac 1{2L}(p_{L}-p_{L-1}).
\ena

Now we rewrite $x p_L$ as
\bea
xp_L&=&\sum_{n=0}^{\infty}\frac{\G(n+L-1-i\e)\G(n-L-2-i\e)}
{\G(n+1-2i\e)n!}x^{n+1}\nonumber\\
&=&\sum_{n=0}^{\infty}\frac{n(n-2i\e)}{(n-L-3-i\e)(n+L-2-i\e)}A_{L,n}x^n
\nonumber\\
&=&\sum_{n=0}^{\infty}\left[1+\frac{(L+3+i\e)(L+3-i\e)}{(2L+1)(n-L-3-i\e)}
 -\frac{(L-2+i\e)(L-2-i\e)}{(2L+1)(n+L-2-i\e)}\right] A_{L,n}x^n.
\nonumber\\
\ena
By using Eqs.(A.4) and (A.5), we obtain 
\bea
xp_{n+\n}&=&{(n+\n+3+i\e)(n+\n+3-i\e)\over{2(n+\n+1)(2n+2\n+1)}}p_{n+\n+1}
\nonumber\\
&+&{1\over{2}}\left[1-{4+\e^2\over{(n+\n)(n+\n+1)}}\right]
p_{n+\n}\nonumber\\
&+&{(n+\n-2+i\e)(n+\n-2-i\e)\over{2(n+\n)(2n+2\n+1)}}p_{n+\n-1}.
\ena

Similarly, we rewrite $x(1-x)p_L'$ as
\bea
x(1-x)p_L'&=&\sum_{n=0}^{\infty}nA_{L,n}(x^n-x^{n+1})\nonumber\\
&=&\sum_{n=0}^{\infty}n\left[1-\frac{(n-1)(n-2i\e)}
{(n-L-3-i\e)(n+L-2-i\e)}\right] A_{L,n}x^n
\nonumber\\
&=&-\sum_{n=0}^{\infty}n\left[ \frac{(L+3-i\e)(L+2+i\e)}{(2L+1)(n-L-3-i\e)}
+\frac{(L-2+i\e)(L-1-i\e)}{(2L+1)(n+L-2-i\e)}\right] A_{L,n}x^n
\nonumber\\
&=&-\sum_{n=0}^{\infty}\left[ 4+\frac{(L+3+i\e)(L+3-i\e)(L+2+i\e)}{2L+1}
\frac 1{n-L-3-i\e}\right. \nonumber\\
&&\left. -\frac{(L-2+i\e)(L-2-i\e)(L-1-i\e)}{2L+1}\frac 1{n+L-2-i\e}
\right] A_{L,n}x^n.
\nonumber\\
\ena
By using Eqs.(A.4) and (A.5), we find 

\bea
x(1-x)p'_{n+\n}&=&-{(n+\n+3+i\e)(n+\n+3-i\t)(n+\n+2+i\e)
\over{2(n+\n+1)(2n+2\n+1)}}p_{n+\n+1}\nonumber\\
&+&{1\over{2}}\left[-2+i\e+{(4+\e^2)(1+i\e)\over{(n+\n)(n+\n+1)}}\right]
p_{n+\n}\nonumber\\
&+&{(n+\n-2+i\e)(n+\n-2-i\e)(n+\n-1-i\e)\over{2(n+\n)(2n+2\n+1)}}p_{n+\n-1}.
\ena 
 
\noindent
(b) Proof of the recurrence  relations among Coulomb wave  functions\\

Let us define 
\bea
B_L&=&e^{-iz}(2z)^{L} z,\hskip 1cm 
C_{L,n}=\frac{\G(n+L+1+i\e)}{\G(n+2L+2)n!},
\ena
then $F_L$ is written by
\bea
F_L=B_L\sum_{n=0}^{\infty}C_{L,n}(2iz)^n.
\ena
Now let us  consider the decomposition of $F_{L+1}$  as 
\bea
F_{L+1}&=&B_L \frac 1{i}\sum_{n=1}^{\infty}\frac
{\G(n+L+1+i\e}{\G(n+2L+3)(n-1)!}(2iz)^n
\nonumber\\
&=&B_L\frac 1{i}\sum_{n=0}^{\infty}\frac{n}{n+2L+2}C_{L,n} (2iz)^n
\nonumber\\
&=&\frac 1{i}\left[ F_L -2(L+1)B_L\sum_{n=0}^{\infty}\frac{C_{L,n}
(2iz)^n}{n+2L+2}\right ].
\ena 
Thus we have 
\bea
B_L\sum_{n=0}^{\infty}\frac{C_{L,n}(2iz)^n}{n+2L+2}=\frac 1{2(L+1)}
(F_L-iF_{L+1}).
\ena
Similarly, 
\bea
F_{L-1}&=&B_L i\sum_{n=0}^{\infty}\frac{\G(n+L+i\e)}{\G(n+2L)(n)!}(2iz)^{n-1}
\nonumber\\
&=&B_L i\left[\frac{\G(L+i\e)}{\G(2L)}\frac1{2iz}+ 
\sum_{n=0}^{\infty}\frac{n+2L+1}{n+1}C_{L,n}(2iz)^{n}\right].
\ena
Then we find
\bea
B_L\sum_{n=0}^{\infty}\left[ \frac{\G(L+i\e)}{\G(2L+1)}\frac 1{2iz}+
\frac{C_{L,n}(2iz)^n}{n+1}
\right ]=-\frac 1{2L}(F_L+iF_{L-1}).
\ena
Now we evaluate $F_L/z$ as 
\bea
\frac{F_L}{z}&=&2iB_L\left[ \sum_{n=0}^{\infty}\frac{\G(n+L+1+i\e)}
{\G(n+2L+2)n!}(2iz)^{n-1}
\right]\nonumber\\
&=&2iB_L\left[ \frac{\G(L+1+i\e)}{\G(2L+2)}\frac 1{2iz}+
\sum_{n=0}^{\infty}\frac{n+L+1+i\e}{(n+2L+2)(n+1)}C_{L,n}(2iz)^n\right]
\nonumber\\
&=&2iB_L\left \{\frac{L+i\e}{2L+1}\left[\frac{\G(L+i\e)}{\G(2L+1)}\frac1{2iz}+
\sum_{n=0}^{\infty}\frac{C_{L,n}(2iz)^n}{n+1}
\right]\right.\nonumber\\
&&\left.+\frac{L+1-i\e}{2L+1}
\sum_{n=0}^{\infty}\frac{1}{n+2L+2}C_{L,n} (2iz)^n\right \}.
\nonumber\\
\ena
By using Eqs.(A.13) and (A.14), we find 
\bea
{1\over{z}}F_{n+\n}&=&{(n+\n+1-i\e)\over{(n+\n+1)(2n+2\n+1)}}F_{n+\n+1}
+{\e\over{(n+\n)(n+\n+1)}}F_{n+\n}\nonumber\\
&+&{(n+\n+i\e)\over{(n+\n)(2n+2\n+1)}}F_{n+\n-1}.
\ena

Similarly, 
\bea
F_L'=-iF_L+\frac{L+1}z F_L + 2iB_L\sum_{n=0}^{\infty}\frac{n+L+1+i\e}{
n+2L+2}C_{L,n}(2iz)^n.
\ena
By using Eqs.(A.13) and (A.14), we obtain 
\bea
F'_{n+\n}&=&-{(n+\n)(n+\n+1-i\e)\over{(n+\n+1)(2n+2\n+1)}}F_{n+\n+1}
+{\e\over{(n+\n)(n+\n+1)}}F_{n+\n}\nonumber\\
&+&{(n+\n+1)(n+\n+i\e)\over{(n+\n)(2n+2\n+1)}}F_{n+\n-1}.
\ena

\apsc{The convergence region of solutions}

From Eq.(2.13), we find 
\bea
\displaystyle \lim_{n \rightarrow \infty} n\frac{a_n^{\n}}{a_{n-1}^{\n}}=
-\lim_{n \rightarrow -\infty}  n\frac{a_n^{\n}}{a_{n+1}^{\n}}
=-\frac {i\e}2.
\ena
By combining the large $n$ behavior of hypergeometric 
functions\cite{Bateman}, we find 
\bea
\displaystyle \lim_{n \rightarrow \infty} 
\frac{n a_n^{\n}p_{n+\n}(x)}{a_{n-1}^{\n}p_{n+\n -1}(x)}=
-\lim_{n \rightarrow -\infty}  
\frac{n a_n^{\n}p_{n+\n}(x)}{a_{n+1}^{\n}p_{n+\n +1}(x)}
=\frac {i\e}{2}[1-2x +((1-2x)^2-1)^{1/2}].
\ena
Thus the hypergeometric series converges in all the 
complex plane of $x$ except for $x=\infty$.

As for the convergence of Coulomb series in Eq.(3.3), we find\cite{Bateman}
\bea
\displaystyle \lim_{n \rightarrow \infty} 
\frac{F_{n+\n}(z)}{nF_{n+\n-1 }(z)}=
\lim_{n \rightarrow -\infty}  
\frac{F_{n+\n}(z)}{nF_{n+\n +1}(z)}
=\frac{2}{z},
\ena
 that 
\bea
\displaystyle \lim_{n \rightarrow \infty} 
\frac{a_n^{\n}F_{n+\n}(z)}{a_{n-1}^{\n}F_{n+\n-1 }(z)}=
\lim_{n \rightarrow -\infty}  
\frac{a_n^{\n}F_{n+\n}(z)}{a_{n+1}^{\n}F_{n+\n +1}(z)}
=-i\frac{\e}{z}.
\ena
Thus, the series converges for $z > \e $ or 
$\mid x\mid  > 1$.

\end{document}